# Compact Microstrip Triplexer Using Square Open-Loop Resonators with High Isolation for 5G Sub-6 GHz Applications

Bhaskarareddy N C
*Dept. of Electrical and Electronic*
*School of Engineering*
*University of Greenwich, London, UK*
nb6621j@gre.ac.uk

Rashmi Ravi
*Dept. of Electrical and Electronic*
*School of Engineering*
*University of Greenwich, London, UK*
rr8438i@gre.ac.uk

Augustine O. Nwajana
*Dept. of Electrical and Electronic*
*School of Engineering*
*University of Greenwich, London,*
a.o.nwajana@greenwich.ac.uk

***Abstract***— *The rapid proliferation of multi-standard fifth-generation (5G) New Radio sub-6 GHz systems have intensified the demand for compact, high-isolation multiplexing components capable of simultaneously routing multiple frequency channels through a single shared antenna port without cross-band interference. This paper presents the design and full-wave electromagnetic (EM) simulation of a compact microstrip triplexer operating at 2.2, 2.6, and 3.0 GHz for 5G sub-6 GHz applications. The proposed triplexer employs square open-loop resonators (SOLRs) arranged as three independent three-pole Chebyshev bandpass filter channels, yielding nine resonator poles in total. Each channel is synthesised from the standard normalised Chebyshev lowpass prototype with element values* $g_0 = g_4 = 1.0$, $g_1 = g_3 = 0.8516$, *and* $g_2 = 1.1032$, *with a fractional bandwidth of 3% per channel and 50 Ω system impedance. The three channels are integrated at a common input port via a T-junction, with connecting transmission line stubs dimensioned to enforce high inter-channel isolation. The triplexer is implemented on Rogers RT/Duroid 6010LM substrate with dielectric constant of 10.7, loss tangent of 0.0023, and thickness of 1.27mm. Full-wave EM simulation results demonstrate return losses of 21.1 dB, 23.1 dB, and 22.8 dB; insertion losses of 1.08 dB, 1.01 dB, and 0.98 dB; and inter-channel isolations of 45.7 dB, 45.2 dB, and 45.7 dB for port pairs* $S_{32}$, $S_{42}$, *and* $S_{43}$, *respectively. The achieved inter-channel isolation exceeding 45 dB represents a significant improvement over the majority of recently reported microstrip triplexer designs.*

***Keywords—****Chebyshev; high isolation; microstrip; resonator; square open-loop; sub-6 GHz; triplexer.*

## I. INTRODUCTION

The ongoing global deployment of fifth generation (5G) New Radio (NR) infrastructure, together with the continued coexistence of Long-Term Evolution (LTE) and Wi-Fi networks, has placed increasingly stringent demands on radio frequency (RF) front-end architectures. Modern transceivers must simultaneously support multiple frequency bands while maintaining minimal inter-channel interference, low signal degradation, and compact form factors appropriate for integration into both portable user equipment and base-station hardware [1], [2]. In this environment, passive multiplexing components capable of routing two or more frequency channels through a single shared antenna port have become critical enabling technologies [3].

Among multiplexing topologies, the triplexer occupies a particularly important position. A triplexer is a four-port passive device comprising three bandpass filter (BPF) channels connected to a common input port, enabling simultaneous and independent operation across three distinct frequency bands with high inter-channel isolation [4]. Compared to cascaded diplexer arrangements, which accumulate insertion loss, add matching-network complexity, and degrade isolation with each additional stage, a monolithic triplexer consolidates all three filtering and routing functions into a single device, thereby reducing circuit complexity, cumulative insertion loss, and physical footprint. These advantages are especially desirable in 5G sub-6 GHz deployments targeting the 2.2 GHz, 2.6 GHz, and 3.0 GHz bands assigned to New Radio operations in multiple international regulatory frameworks [5].

Considerable research effort has been directed toward microstrip triplexer design owing to the planar geometry, low fabrication cost, and compatibility with standard printed circuit board processes [2]. The published literature reveals several distinct resonator technology families, each with characteristic performance trade-offs.

Stepped-impedance resonators (SIRs) represent one of the earliest and most widely studied approaches. By adjusting the impedance ratio of alternating high- and low-impedance sections, SIRs achieve harmonic suppression and size reduction and have been employed in triplexer designs with acceptable insertion loss [6]. However, the reliance on external matching circuits between filter channels introduces additional loss and complexity, and inter-channel isolation typically remains below 25 dB for standard SIR topologies [4].

Complementary split-ring resonator (CSRR) loaded coupled-line triplexers offer compactness through the transmission-zero-enhancing properties of CSRR loading. The design reported in [7] demonstrates a CSRR-loaded triplexer at 1.4/1.8/3.2 GHz achieving a normalised footprint of $0.07\lambda g^2$; however, inter-channel isolation of approximately 20 dB limits suitability for demanding multi-standard front-ends. Coupled-line and stepped-impedance cell triplexers [8] achieve low insertion loss (0.63–0.81 dB) and good return loss (24–24.7 dB), but isolation is commonly not reported.

Meandered loop resonators [9] and patch-and-spiral cell configurations [10] have been reported as compact solutions, with [10] achieving $0.017\lambda g^2$ at 1.9/2.5/3.35 GHz. Although these designs achieve useful miniaturisation, isolation levels remain in the 20–25 dB range, insufficient for applications where channel crosstalk must be rigorously controlled. Among sub-6 GHz designs, those explicitly achieving isolation above 40 dB remain comparatively rare in the published literature [5], underscoring the need for topology-level isolation solutions rather than post-design filtering.

The square open-loop resonator (SOLR) is a well-established building block for coupled-resonator bandpass filter design [11]. The SOLR is formed by folding a half-wavelength (λg/2) straight microstrip resonator into a square loop with a coupling gap at one end, yielding a side length of λg/8, a fourfold reduction in linear dimension relative to the straight resonator. On the high-permittivity Rogers RT/Duroid 6010LM substrate (εr = 10.7), the guided wavelength is further compressed, yielding physically compact resonator dimensions advantageous for multi-channel devices. Despite its widespread use in filter and diplexer design [12], the application of SOLRs to triplexer architectures on high-permittivity substrates has received limited systematic attention [13].

This paper presents a compact microstrip triplexer designed using three independent three-pole Chebyshev BPF channels, each realised with SOLRs, operating simultaneously at 2.2 GHz, 2.6 GHz, and 3.0 GHz on RT/Duroid 6010LM substrate. The three filter channels are integrated at a common input port through a T-junction, the connecting stubs of which are dimensioned to enforce physical separation of the channel paths and achieve inter-channel isolation significantly exceeding that of most reported microstrip triplexers operating in the sub-6 GHz range.

The remainder of this paper is organised as follows. Section II describes the Chebyshev prototype filter synthesis and theoretical circuit configuration. Section III details the SOLR-based microstrip layout and practical design methodology. Section IV presents and analyses the EM simulation results together with a comparative performance analysis against published works. The Conclusion summarises the contribution and outlines the ongoing experimental validation.

## II. THEORETICAL CIRCUIT CONFIGURATION

### *A. Resonator Transformation and Size Reduction*

The size reduction principle employed in this work is illustrated in Fig. 1. A conventional half-wavelength (λg/2) straight microstrip resonator is folded into a square open-loop configuration with dimensions λg/8 × λg/8, achieved by bending the resonator at three points to form a near-closed square loop with a coupling gap at one end. This folding reduces the resonator side length to one-eighth of the guided wavelength at the operating frequency, yielding physically compact dimensions on the high-permittivity RT/Duroid 6010LM substrate. The standard SOLR coupling geometry is well characterised [1] and enables confident mapping from coupling coefficient to physical gap dimension for each of the nine resonators in the complete device.

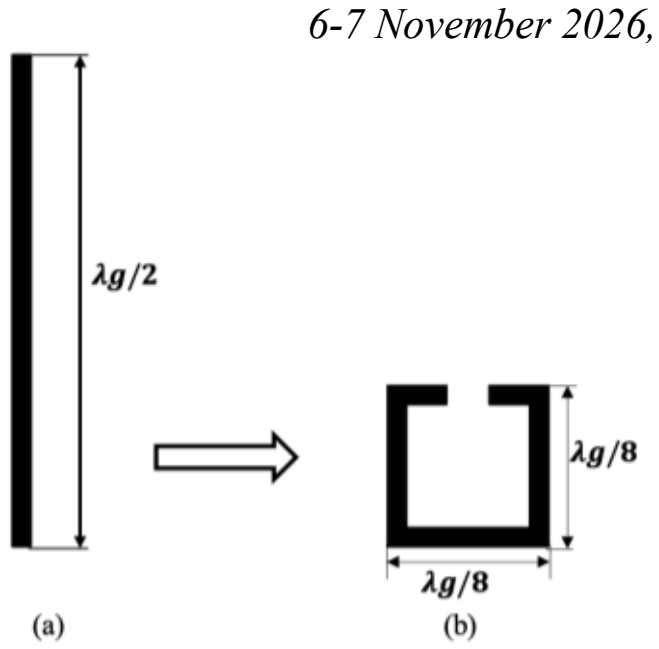

*Fig. 1. Resonator transformation: (a) half-wavelength (λg/2) straight microstrip resonator folded into (b) a compact square open-loop resonator (SOLR) of dimensions λg/8 × λg/8.*

### *B. Chebyshev Filter Prototype Synthesis*

The circuit configuration for the proposed triplexer is achieved by first synthesising three identical three-pole Chebyshev bandpass filter channels from the standard normalised Chebyshev lowpass prototype filter element-values of $g_0 = g_4 = 1.0$, $g_1 = g_3 = 0.8516$, and $g_2 = 1.1032$. These values correspond to a three-pole equal-ripple Chebyshev response, chosen to balance in-band ripple, out-of-band rejection selectivity, and physical complexity. The same prototype values are employed in the design of the filtering power divider reported in [14], [15], confirming their suitability for coupled-resonator designs on RT/Duroid 6010LM.

The design centre frequencies for the three BPF channels are 2.2 GHz, 2.6 GHz, and 3.0 GHz, corresponding to key 5G New Radio sub-6 GHz frequency allocations. Each channel is assigned a fractional bandwidth (FBW) of 3% of its respective centre frequency, with input/output characteristic impedance $Z_0 = 50\ \Omega$ throughout. The three channels are denoted $BPF_{2.2}$, $BPF_{2.6}$, and $BPF_{3.0}$.

### *C. Coupling Parameters*

The inter-resonator coupling coefficient M and external quality factor Qe for each filter channel are determined from the standard coupled-resonator filter design formulations [1], [2]. The coupling coefficient M governs the in-band ripple and bandwidth of each channel and is given by (1):

$$M = \frac{FBW}{\sqrt{g_1 g_2}} = 0.0309 \qquad (1)$$

The external quality factor Qe governs the coupling between the input port and the first resonator of each channel, and between the last resonator and its respective output port, and is given by (2):

$$Q_e = \frac{g_0 g_1}{FBW} = 28.387 \qquad (2)$$

Since M and Qe depend only on the Chebyshev prototype values and FBW, and not on the centre frequency, they are identical for all three filter channels. This design symmetry simplifies the layout, as the same coupling coefficient (M = 0.0309) and the same external quality factor (Qe = 28.387) apply to each channel, realised by physical gap and tapping dimensions scaled to the resonator size at each operating frequency.

### *D. Circuit Model*

The three BPF channels are connected to the common input port through a T-junction via transmission line stubs $TL_1$, $TL_2$, and $TL_3$, as shown in Fig. 2(a). Each channel consists of three LC resonators coupled by admittance inverters $J_{01}$ and $J_{12}$, which represent the input/output coupling and the inter-resonator coupling, respectively. The resonator and coupling element values for each channel are summarised in the caption of Fig. 2(a). The element values decrease monotonically from the 2.2 GHz channel to the 3.0 GHz channel, consistent with the inverse relationship between resonator capacitance and operating frequency at fixed impedance. The theoretical responses showing well-defined passbands at all three centre frequencies, with high adjacent-channel selectivity and low inter-channel leakage, are presented in Fig. 2(b), confirming the correctness of the synthesis before proceeding to physical layout.

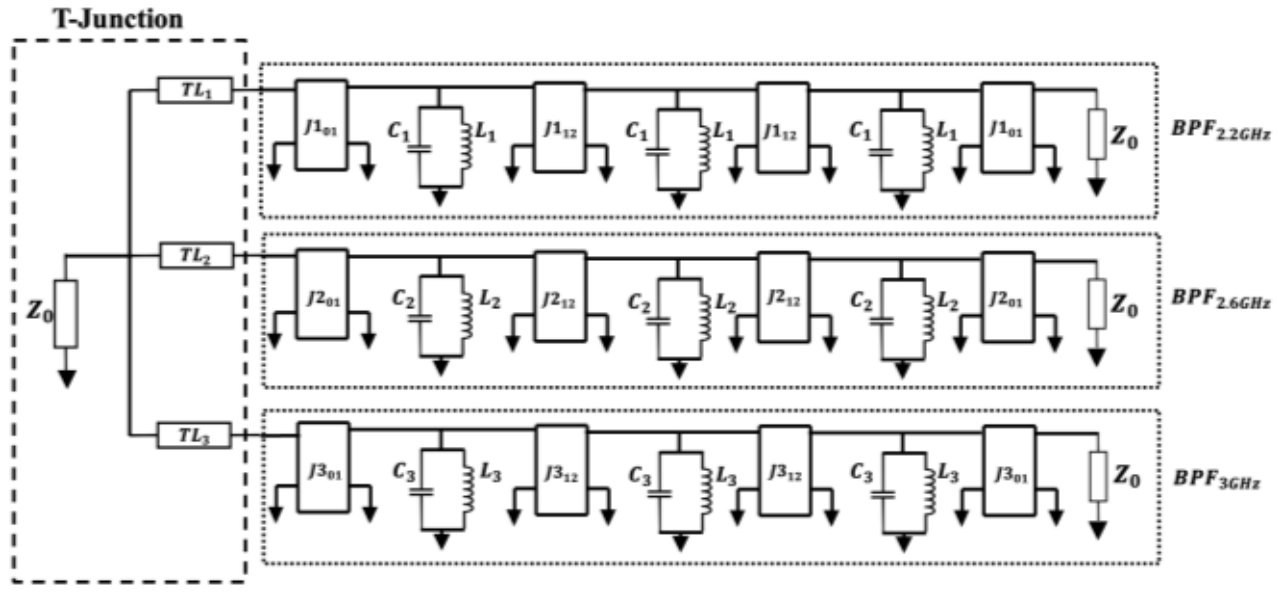

*Fig. 2(a). Theoretical circuit arrangement comprising a T-junction and three three-pole Chebyshev BPF channels 2.2 GHz, 2.6 GHz, and 3.0 GHz ($Z_0$ = 50 Ω; $BPF_{2.2}$GHz: $J1_{01}$=1.4468 pF, $J1_{12}$=1.266 pF, $C_1$=41.071 pF, $L_1$=0.1274 nH; $BPF_{2.6}$GHz: $J2_{01}$=1.2242 pF, $J2_{12}$=1.0712 pF, $C_2$=34.752 pF, $L_2$=0.1078nH; $BPF_{3.0}$GHz: $J3_{01}$=1.0610 pF, $J3_{12}$=0.9284 pF, $C_3$=30.119 pF, $L_3$=0.09344 nH)*

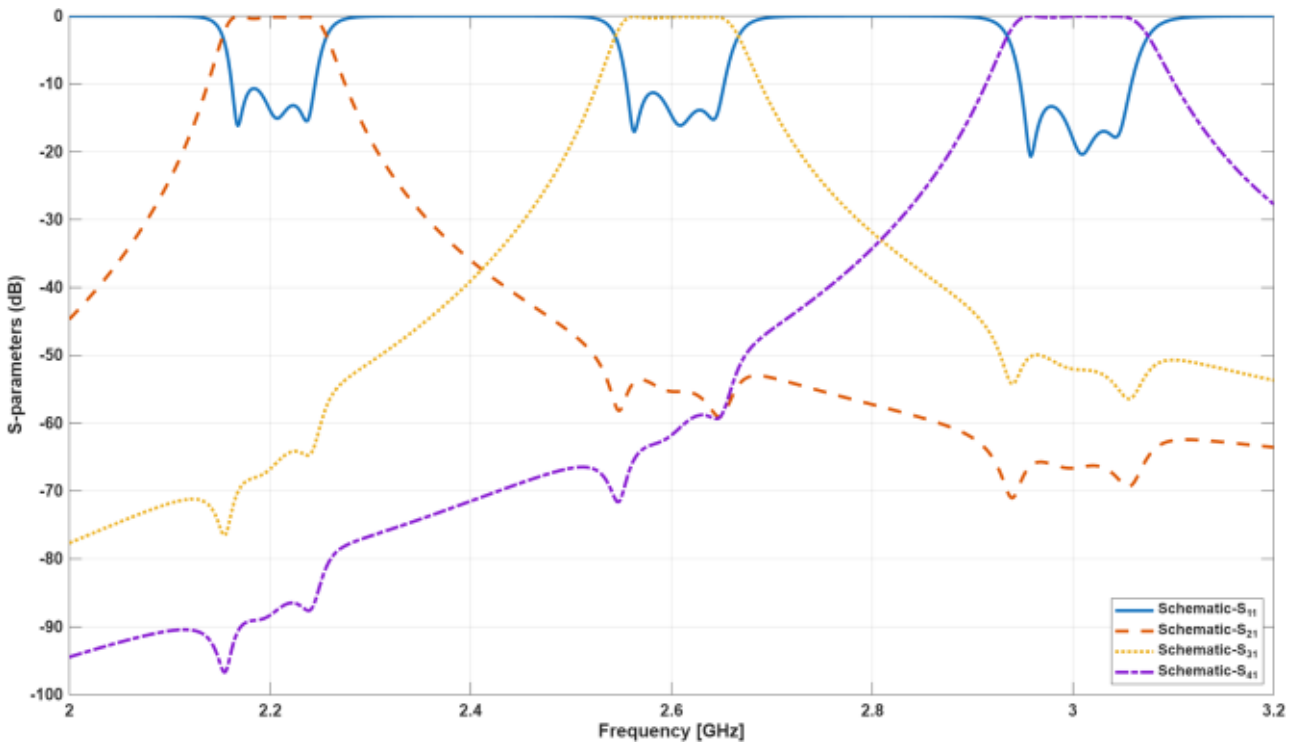

*Fig. 2(b). Theoretical simulation responses of the proposed triplexer.*

## III. MICROSTRIP LAYOUT AND PRACTICAL ARRANGEMENT

### *A. SOLR Physical Dimensions*

The triplexer design and full-wave simulation were carried out using microstrip technology with the SOLR as the fundamental building block. The guided wavelength λg and microstrip line width w for each resonator are determined from the substrate parameters and the respective resonant frequency using the standard quasi-TEM microstrip design equations [1]. For a 50 Ω characteristic impedance on Rogers RT/Duroid 6010LM (εr = 10.7, h = 1.27 mm), the microstrip line width is approximately 1.14 mm, corresponding to a width-to-height ratio w/h ≈ 0.87 and an effective permittivity εr,eff ≈ 7.11, yielding a guided wavelength of approximately 43.3 mm at 2.6 GHz.

Each SOLR is designed to resonate at the circuit-model specified centre frequency of its respective filter channel: 2.2 GHz, 2.6 GHz, and 3.0 GHz. The physical side length of each SOLR is λg/8 at the respective operating frequency, yielding three distinct but geometrically similar resonator dimensions. The inter-resonator coupling gaps are adjusted to realise M = 0.0309, and the feedline tapping position is adjusted to realise Qe = 28.387, using the coupling characterisation method of Hong and Lancaster [16].

### *B. Coupling Arrangement and T-Junction Integration*

The coupling arrangement for the proposed triplexer is shown in Fig. 3. Each of the three filter channels, $BPF_{2.2}$, $BPF_{2.6}$, and $BPF_{3.0}$, is realised using three SOLRs (resonators A1–A3 for the 2.2 GHz channel, B1–B3 for 2.6 GHz, and C1–C3 for 3.0 GHz), yielding nine resonator poles across the complete device. Port 1 is the common input, with Ports 2, 3, and 4 being the channel outputs of $BPF_{2.2}$, $BPF_{2.6}$, and $BPF_{3.0}$, respectively.

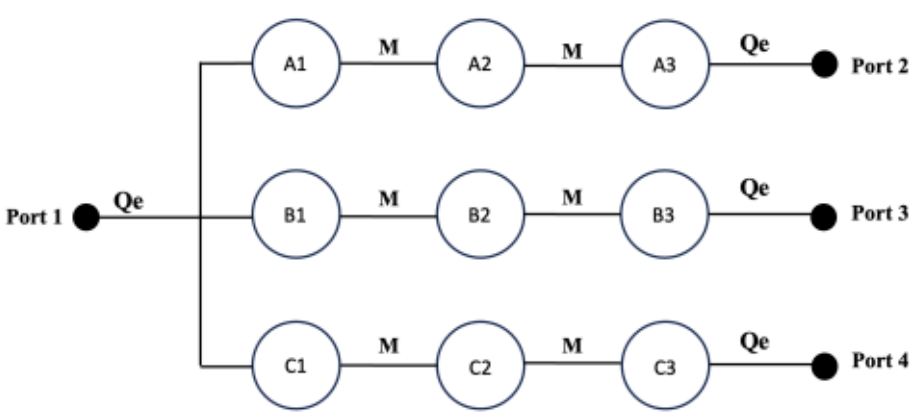

*Fig. 3. Coupling arrangement for the proposed microstrip triplexer. Channel A (A1–A3); Channel B (B1–B3); Channel C (C1–C3). M = inter-resonator coupling coefficient; Qe = external quality factor.*

The three filter channels are connected to the common input port through a T-junction via transmission line stubs $TL_1$, $TL_2$, and $TL_3$. The stub lengths are selected to be approximately quarter-wavelength (λg/4) at the respective channel centre frequencies, presenting a high-impedance condition at the T-junction node for all channels except the intended one, thereby maximising inter-channel isolation. The complete layout was constructed and simulated using Keysight PathWave Advanced Design System (ADS) Momentum EM simulation software on Rogers RT/Duroid 6010LM substrate (εr = 10.7, h = 1.27 mm, tan δ = 0.0023).

### *C. Layout Dimensions and Footprint*

The physical dimensions of the triplexer microstrip layout, with all critical dimensions annotated in millimetres, are shown in Fig. 4(a). The complete active circuit occupies a bounding box of 69.45 mm × 39.7 mm on the substrate.

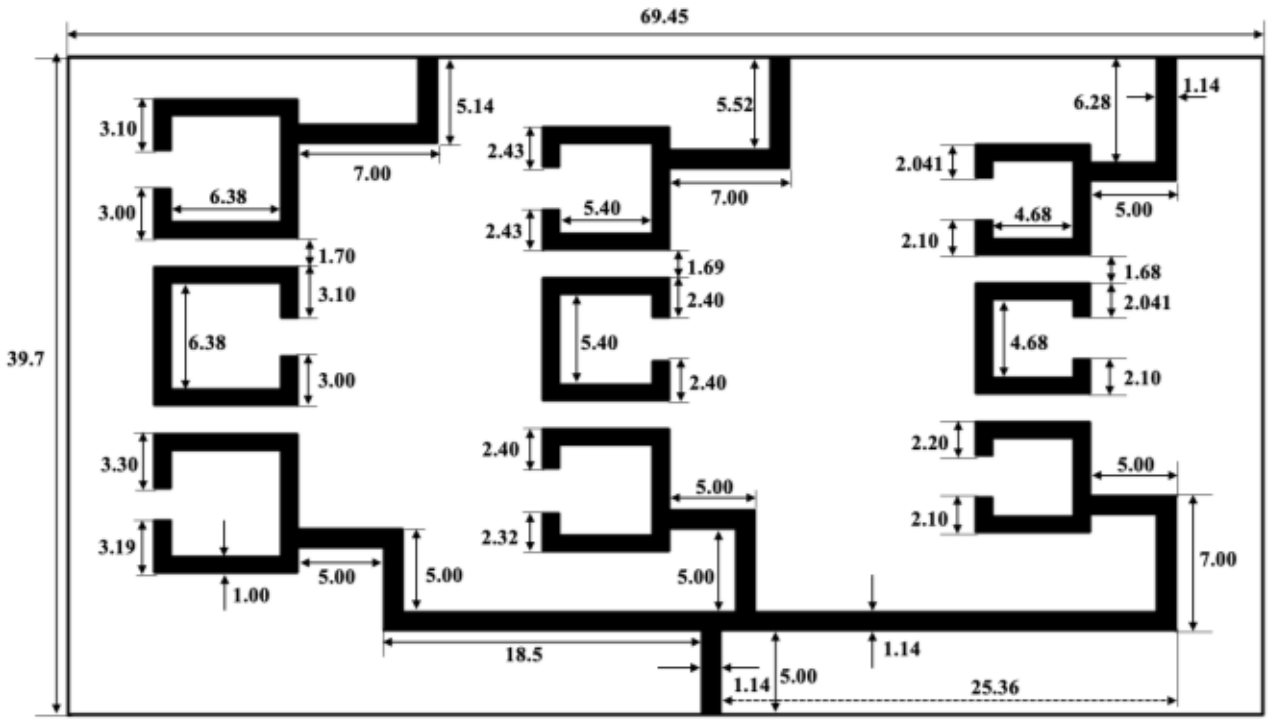

*Fig. 4(a). EM simulation layout of the proposed microstrip triplexer with dimensions in mm.*

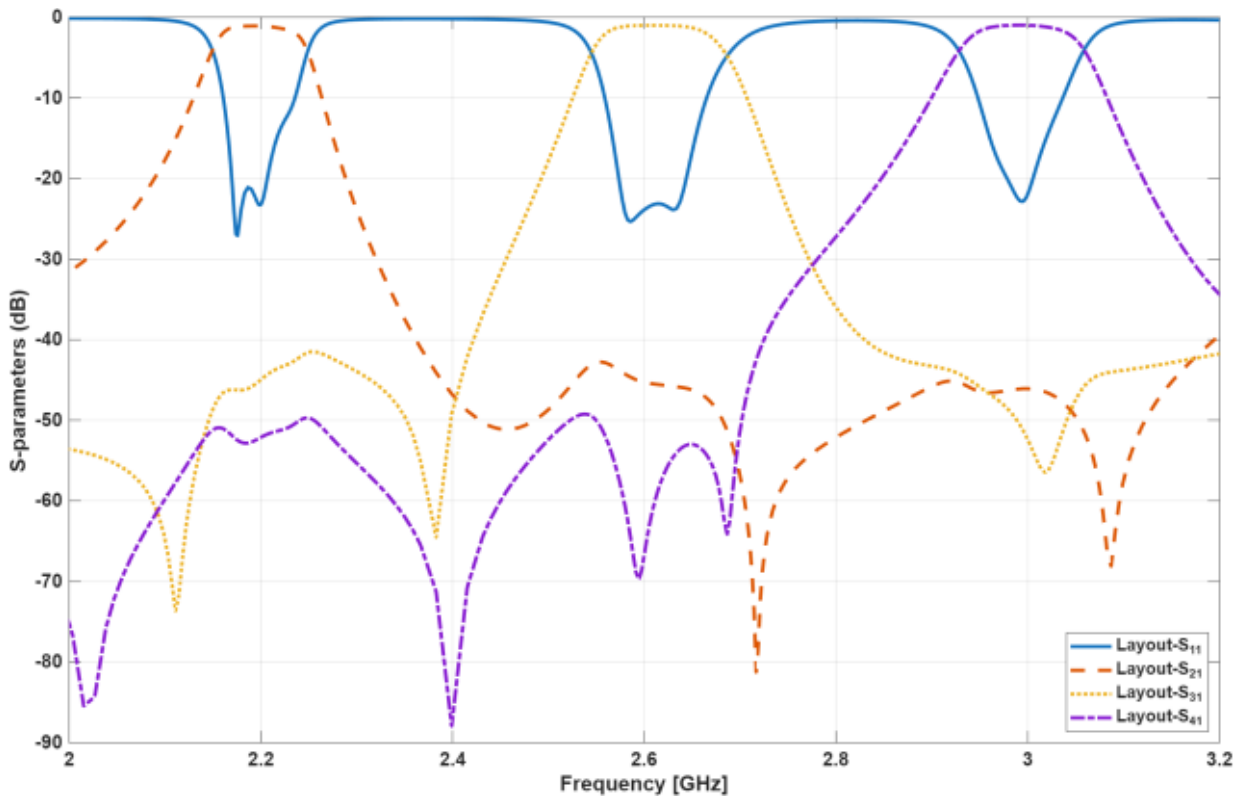

*Fig. 4(b). Full-wave EM simulation S-parameter responses of the proposed triplexer.*

## IV. RESULTS ANALYSIS AND DISCUSSION

### A. *Theoretical-to-EM Simulation Agreement*

This section presents and analyses the theoretical circuit simulation and full-wave EM simulation results for the proposed microstrip triplexer. The results are jointly presented in Fig. 5(a) for ease of analysis and comparison. The solid-line plots represent the theoretical circuit model responses, while the dashed-line plots denote the full-wave EM simulation results obtained using Keysight PathWave ADS Momentum. Good agreement is observed between the theoretical and EM simulation responses across all three channels.

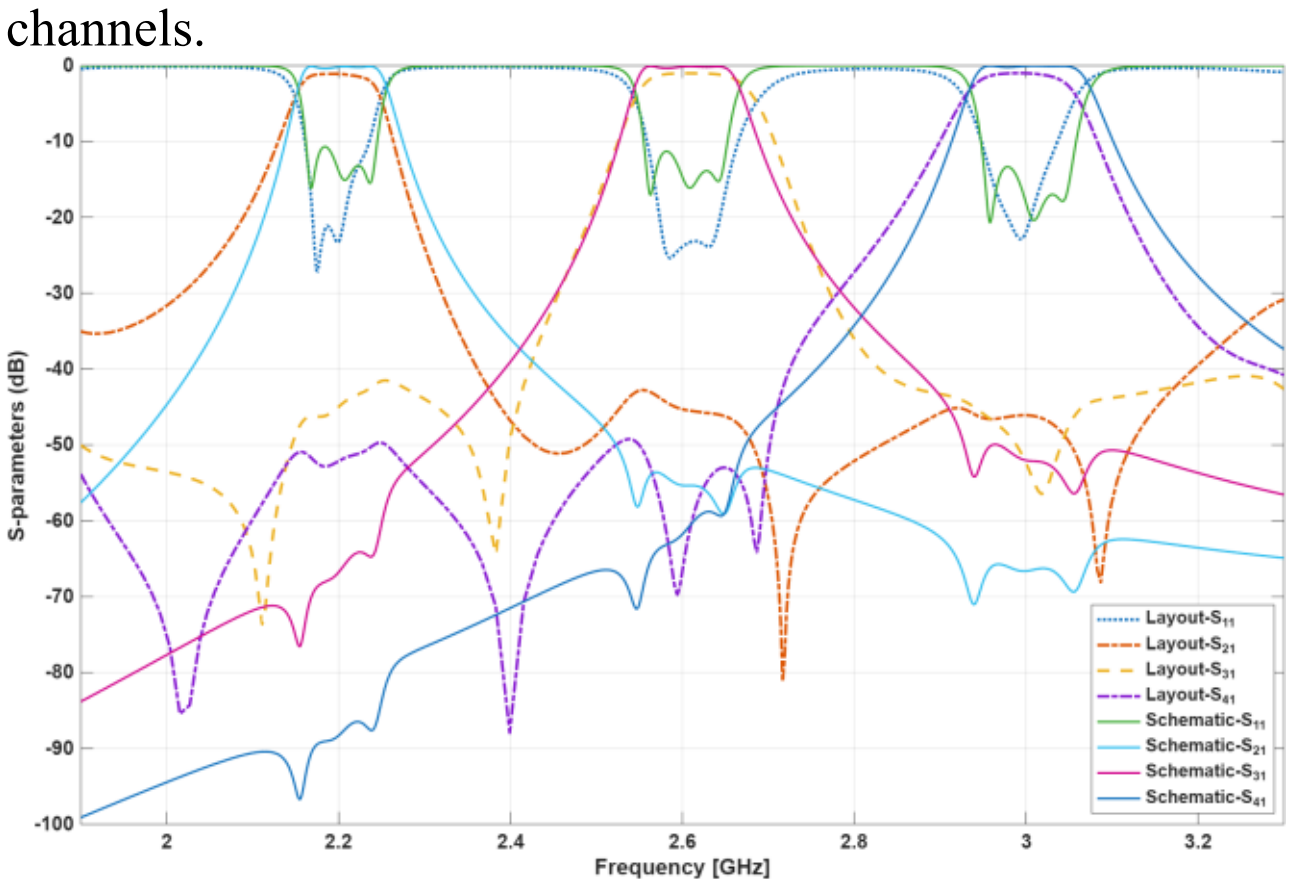

*Fig. 5(a). Theoretical circuit model and EM simulation results comparison for the proposed microstrip triplexer.*

### B. *S-Parameter Performance*

From the S-parameter responses of Fig. 5(a), the proposed triplexer transmits at the three specified centre frequencies of 2.2 GHz, 2.6 GHz, and 3.0 GHz with EM simulation return losses of 21.1 dB, 23.1 dB, and 22.8 dB, respectively. These values comfortably exceed the 20 dB threshold generally regarded as the benchmark for well-matched microwave devices. The corresponding EM simulation insertion losses are 1.08 dB, 1.01 dB, and 0.98 dB at the respective channel centre frequencies, all below 1.1 dB, representing low-loss performance for a nine-pole device.

It is important to note that the reported EM simulation results assume a copper conductor with a cladding thickness of 35 µm and a conductivity of $5.8 \times 10^7$ S/m. Substrate thickness variation and metal surface roughness were not incorporated into the simulation model. These factors are expected to introduce a modest additional insertion loss in the fabricated prototype, consistent with the observations reported in [17], where measured insertion loss exceeded the simulated value by 0.2–0.5 dB under equivalent substrate and conductor conditions.

### C. *Inter-Channel Isolation*

The isolation responses of the proposed triplexer are shown in Fig. 5(b). The EM simulation isolation values, representing the transmission between channel output ports with the common input terminated, are 45.7 dB for $S_{32}$ (between the 2.2 GHz and 2.6 GHz channel outputs), 45.2 dB for $S_{42}$ (between the 2.2 GHz and 3.0 GHz channel outputs), and 45.7 dB for $S_{43}$ (between the 2.6 GHz and 3.0 GHz channel outputs). All three port-pair isolations exceed 45 dB across the relevant operating bands.

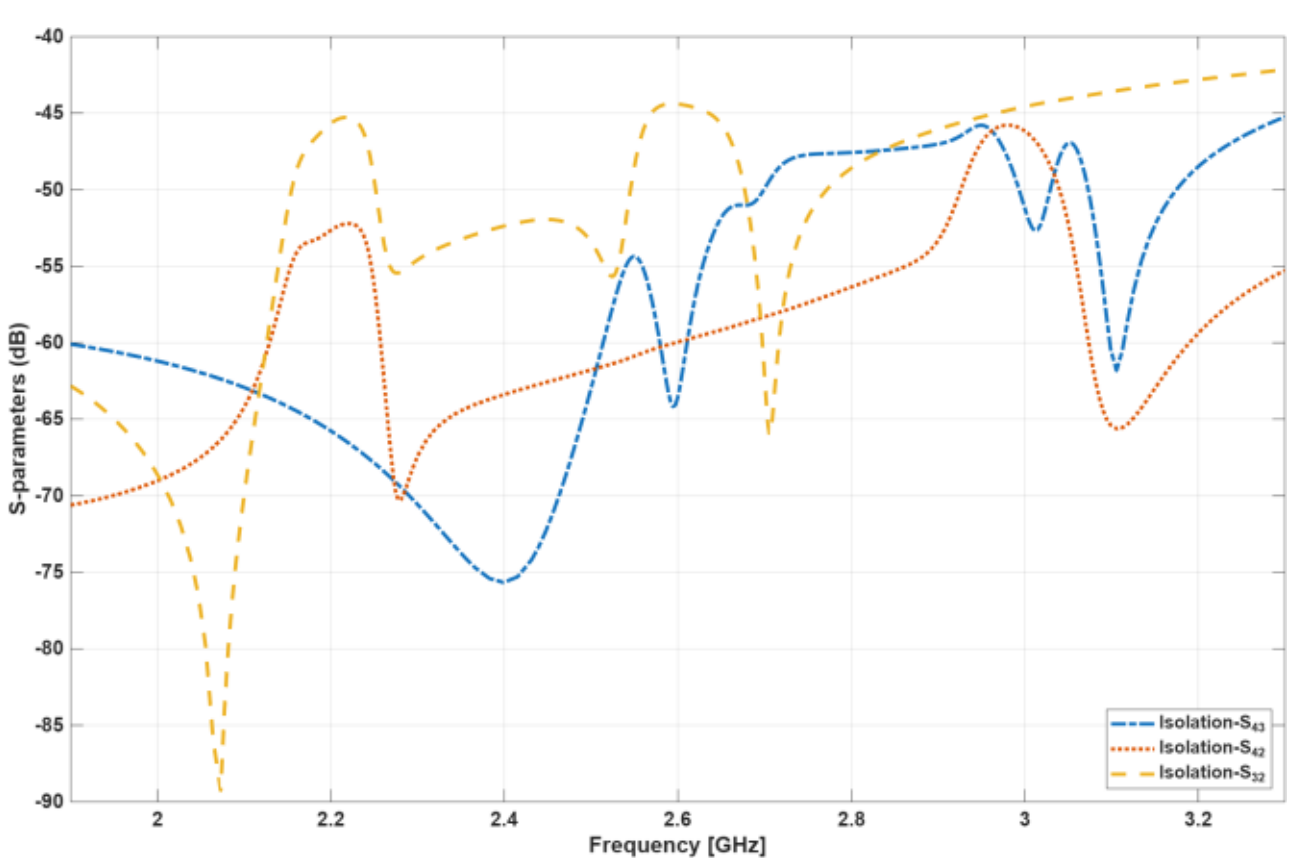

*Fig. 5(b). Isolation responses of the proposed microstrip triplexer.*

This high isolation performance is directly attributed to the T-junction integration scheme employed in connecting the three filter channels at the common input port. The T-junction stubs, dimensioned to approximately λg/4 at each channel frequency, present a high-impedance condition to all channels other than the intended one at the junction node, suppressing inter-channel signal leakage. The three-pole Chebyshev filter order further contributes to isolation by ensuring that each channel presents high rejection at the passband frequencies of the adjacent channels. The combined effect of the junction-level isolation enforcement and the filter-order selectivity yields isolation levels substantially exceeding the 20–25 dB range commonly reported for microstrip triplexers based on coupled-line, SIR, or CSRR topologies [6]–[10].

### *D. Performance Comparison*

The performance of the proposed triplexer is compared against representative state-of-the-art microstrip triplexers in Table I. The comparison covers operating frequencies, filter order, normalised circuit size, transmission line technology, return loss, insertion loss, and inter-channel isolation. The isolation column is included specifically to contextualise the principal contribution of this work, as it is absent from most published triplexer comparisons in the literature.

As evident from Table I, the proposed design achieves the highest reported inter-channel isolation (> 45.2 dB) among microstrip triplexers operating in the sub-6 GHz frequency range, with the sole exception of the design in [13], which targets substantially higher frequencies (5.0/16.6/42 GHz) using a specialised miniaturised architecture. Among designs operating in the 2–4 GHz range directly comparable to the proposed work, the next highest reported isolation is > 40 dB in [5], achieved using a hybrid lowpass–bandpass channel topology that constrains the achievable passband shape. The proposed SOLR-based design achieves comparable isolation using three uniform bandpass channels of equal filter order, providing a symmetric, controllable, and reproducible frequency response. The insertion loss of 0.98–1.08 dB is competitive for a nine-pole three-channel device on RT/Duroid 6010LM, and the return loss of 21.1–23.1 dB exceeds the 20 dB benchmark across all three channels.

*Table I. Performance Comparison with Related Literature*

| *Ref.* | *$f_0$ (GHz)* | *Filter order* | *TL[a]* | *RL[b] (dB)* | *IL[c] (dB)* | *Isolation (dB)* |
|---|---|---|---|---|---|---|
| [5] | 1.0/2.4/5.8 | 3 | MS | — | <1.5 | ~40 |
| [6] | Varies | 3 | MS | — | <2.0 | -20 |
| [7] | 1.4/1.8/3.2 | 3 | MS | >20 | <2.0 | ~20 |
| [8] | 2.67/3.1/3.43 | 3 | MS | 24.5/24/24.7 | 0.72/0.63/0.81 | — |
| [9] | 2.5/2.8/3.5 | 3 | MS | — | <3.0 | >20 |
| [10] | 1.9/2.5/3.35 | 3 | MS | 45/54/40 | 0.25/0.40/0.11 | >20 |
| This work | 2.2/2.6/3.0 | 3 | MS | 21.1/23.1/22.8 | 1.08/1.01/0.98 | 45.7/45.2/45.7 |

[a] transmission line; [b] return loss; [c] insertion loss. MS = microstrip.

## V. CONCLUSION

This paper has presented the design and full-wave electromagnetic simulation of a compact microstrip triplexer using square open-loop resonators for 5G sub-6 GHz multi-standard wireless communication applications. The principal performance achievement is an inter-channel isolation exceeding 45.2 dB across all three port pairs, a level that substantially outperforms the majority of recently reported microstrip triplexers operating in the sub-6 GHz range, as confirmed by the comparative analysis of Table I. This isolation performance is attributed to the T-junction integration scheme employed at the common input port, in which the connecting stubs enforce electromagnetic separation of the three filter channel paths without the need for additional isolation resistors or auxiliary coupling structures.

The proposed triplexer operates simultaneously at the 2.2 GHz, 2.6 GHz, and 3.0 GHz bands, directly aligned with 5G NR sub-6 GHz allocations, with EM simulation return losses of 21.1 dB, 23.1 dB, and 22.8 dB, and insertion losses of 1.08 dB, 1.01 dB, and 0.98 dB across the respective channels. All three channels employ identical three-pole Chebyshev bandpass filter synthesis from the prototype element-values $g_0 = g_4 = 1.0$, $g_1 = g_3 = 0.8516$, $g_2 = 1.1032$, with 3% FBW per channel, yielding nine resonator poles in total and ensuring symmetric, predictable frequency responses. The design is implemented on Rogers RT/Duroid 6010LM substrate ($\varepsilon r = 10.7$, h = 1.27 mm, tan $\delta = 0.0023$).

The good agreement between the theoretical circuit model and the full-wave EM simulation results confirms the validity of the Chebyshev prototype synthesis methodology and the T-junction integration approach employed in this work.